\documentclass{article}
\usepackage{spconf,amsmath,graphicx}
\usepackage{cite}
\usepackage{epstopdf}
\usepackage{amssymb}
\usepackage{caption}
\usepackage{subfigure}
\usepackage{tabularx}
\usepackage{booktabs}
\usepackage{graphicx}
\usepackage{algorithm}
\usepackage{algorithmic}
\usepackage{multirow}
\usepackage{stfloats}
\usepackage{setspace}
\usepackage{microtype}
\title{Efficient Multi-Channel Speech Enhancement with Spherical Harmonics Injection for Directional Encoding}

\name{Jiahui Pan, Pengjie Shen, Hui Zhang, Xueliang Zhang}

\address{
College of Computer Science, Inner Mongolia University, China\\
\texttt{panjiahui@mail.imu.edu.cn, \{cszh,cszxl\}@imu.edu.cn}
}

\begin{document}
\setstretch{0.86}
\small
\maketitle
\begin{abstract}
Multi-channel speech enhancement extracts speech using multiple microphones that capture spatial cues. Effectively utilizing directional information is key for multi-channel enhancement. Deep learning shows great potential on multi-channel speech enhancement and often takes short-time Fourier Transform (STFT) as inputs directly. To fully leverage the spatial information, we introduce a method using spherical harmonics transform (SHT) coefficients as auxiliary model inputs. These coefficients concisely represent spatial distributions. Specifically, our model has two encoders, one for the STFT and another for the SHT. By fusing both encoders in the decoder to estimate the enhanced STFT, we effectively incorporate spatial context. Evaluations on TIMIT under varying noise and reverberation show our model outperforms established benchmarks. Remarkably, this is achieved with fewer computations and parameters. By leveraging spherical harmonics to incorporate directional cues, our model efficiently improves the performance of the multi-channel speech enhancement.

\end{abstract}
\section{Introduction}
\label{sec:intro}
Multi-channel speech enhancement involves extracting a desired speech signal from noisy environments using data captured by multiple microphones. This technique is critical for applications including, but not limited to, video conferencing systems \cite{mehrotra2011realistic,rao2021conferencingspeech}, distant communication \cite{nguyen2014teleimmersive,tan2019real}, and hearing aids \cite{doclo2010acoustic,nossier2019enhanced}. Unlike single-channel methods that rely solely on spectral and temporal properties, multi-channel enhancement uniquely capitalizes on spatial information. By exploiting spatial cues like inter-channel differences, multi-channel systems can substantially improve speech clarity, background noise reduction, and overall listening experience compared to single-channel techniques. However, effectively integrating and processing spatial cues remains an open challenge.

Traditional approaches include spatial filtering methods such as delay-and-sum beamformer \cite{klemm2008improved}, minimum variance distortionless response (MVDR) \cite{souden2009optimal} beamformer, super-directivity beamformer \cite{bitzer2001superdirective}, and others.
These leverage phase and timing differences between microphones to preferentially extract signals from certain directions.
Although these approaches can perform well, their performance depends on reliable estimation of spatial information, which can be challenging to accurately estimate in noisy conditions.

Recently, deep learning has achieved great progress in multi-channel speech enhancement. Earlier deep learning methods for multi-channel enhancement such as Tan et al.'s GCRN \cite{tan2019learning}, Le et al.'s DPCRN \cite{le2021dpcrn} focused on spectral mapping, processing each channel independently.
To better preserve spatial cues, Liu et al. \cite{liu2021inplace} proposed the inplace gated convolutional recurrent neural network (IGCRN) which efficiently retains spatial information in each frequency bin without the downsampling and upsampling used in conventional CRNs.
Later, Tan et al. \cite{tan2022neural} introduced the concept of neural spectro-spatial filtering which jointly optimizes spectral and spatial filtering using a convolutional neural network with densely-connected blocks. This achieves significant gains over prior approaches for multi-microphone speech enhancement.
More recent methods combine DNNs with traditional beamforming to better utilize spatial information. Examples like FasNet \cite{luo2019fasnet}, EabNet \cite{li2022embedding}, and MIMO-Unet \cite{ren2021causal} exploit complementary strengths of deep learning and array processing for state-of-the-art performance. 
However, most of these deep learning methods for multi-channel speech enhancement directly concatenate the STFT from each microphone as the model inputs.  
They rely on the powerful modeling ability of neural network to exploit the spatial information of sound sources. However, traditional STFT representation is difficult to express the spatial information of the sound sources. 
Effectively incorporating spatial information remains an open challenge. Independent per-channel processing fails to capture inter-channel dependencies and spatial relationships that provide valuable context. However, directly modeling full multi-channel spatial correlations is computationally infeasible. More efficient spatial feature extraction is required to incorporate spatial information without excessive complexity and fully capitalize on spatial diversity in multi-channel scenarios.

Fortunately, spherical harmonic coefficients(SHCs) obtained via SHT provide a comprehensive spatial representation of soundfields \cite{kumar2016near,varanasi2020deep}. This spherical harmonic representation offers two key advantages for multi-channel speech enhancement:
\textbf{Effective capture of spatial information \cite{lugasi2020speech}}: Unlike the STFT, SHT primarily captures the spatial distribution characteristics of soundfields. Grounded in spherical harmonics theory, the SHT discerns the spatial attributes of signals arriving from various directions and their inter-relations across microphone channels. Such spatial capture is crucial for multi-microphone speech enhancement. In a multi-microphone array, this transform adeptly captures the spatial orientation of sounds, enabling more precise differentiation between target speech and background noise.
\textbf{Enhanced spatial resolution \cite{rafaely2015fundamentals}}: Spherical harmonics constitute a complete basis for functions defined on the spherical surface. Thus, any spherical function can be represented precisely as a linear combination of spherical harmonics. The SHT facilitates an accurate depiction of a sound field's spatial distribution. The order of the spherical harmonics determines the granularity of spatial feature capture. While lower orders delineate broad spatial patterns, higher orders characterize finer spatial nuances. By selecting a suitable order, the desired spatial resolution can be achieved.
If the spatial information in these SHCs can be fully utilized, this may help compensate for the lack of descriptive spatial modeling in current mainstream approaches. Exploiting this could greatly improve the performance and robustness of multi-channel speech enhancement.

In this paper, we propose a method to fully leverage spatial information using SHCs as auxiliary model inputs. These coefficients concisely represent spatial distributions. Specifically, our model has two encoders, one for the STFT and another for the SHT. By fusing both encoders in the decoder to estimate the enhanced STFT, we effectively incorporate spatial context.
The key contributions are:
\begin{itemize}
    \item The paper innovatively utilizes SHCs as a means to explore and incorporate spatial information for multi-channel speech enhancement.
    \item A unique dual-encoder framework is introduced that combines STFT and SHT processing to enable optimized handling of both spectral-temporal and spatial data.
    \item The model demonstrates superior performance on TIMIT datasets under diverse conditions, outperforming benchmarks with fewer computations and parameters.
\end{itemize}

\section{System Model}
\begin{figure}
    \centering
    \includegraphics[width=0.4\textwidth]{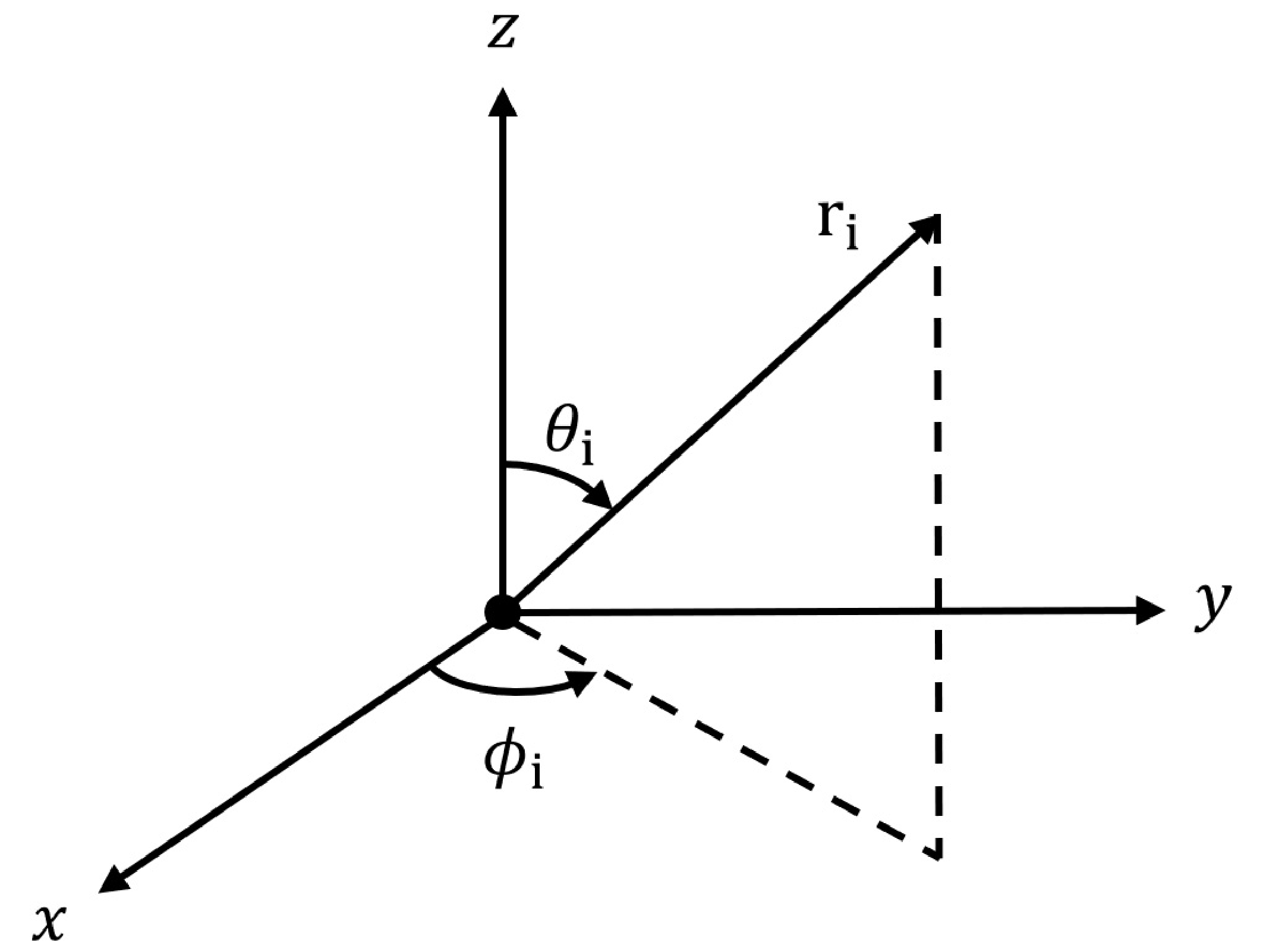}
    \caption{Defined spherical coordinate system.}
    \label{fig:fig1}
\end{figure}

We consider a set of microphone arrays of arbitrary configuration located at the origin of the Cartesian coordinates and composed of $I$ omnidirectional microphones. Let $r_i=\left(r_i \cos \phi_i \sin \theta_i, r_i \sin \phi_i \sin \theta_i, r_i \cos \theta_i\right)^T$ denote the position of the $i$-th microphone of the array, where $r_i$ represents the distance of the $i$-th microphone to the center of the array. The azimuth $\phi_i$ is measured counterclockwise from the x-axis, and the elevation angle $\theta_i$ is measured downward from the z-axis. 
The adopted spherical coordinate system is illustrated in Fig.\ref{fig:fig1}.
The array is assumed to be positioned in a reverberant sound field. According to the image method \cite{Alien1976Image}, the sound pressure in a reverberant environment, generated by a single source in the far field, can be modeled as a sum of $L$ significant plane waves produced by $L$ image sources under free-field conditions. This can be assumed equivalently as $L$ far-field sound sources generating plane waves propagating through space and picked up by the microphones. Let $\Psi_l=\left(\theta_l, \phi_l\right)$ denote the direction of propagation of the $l$-th sound source, and let $k_l=-\left(k \cos \phi_l \sin \theta_l, k \sin \phi_l \sin \theta_l, k \cos \theta_l\right)^T$ represent the wave number vector of the $l$-th plane wave.

\subsection{Space Domain System Model}
The signal received by the $i$-th microphone in the frequency domain can then be expressed as:
\begin{equation}
p_i(k)=\sum_{l=1}^L v_i\left(k, \Psi_l\right) s_l(k)+n_i(k),
\label{p_i(k)}
\end{equation}
where $v_i\left(k, \Psi_l\right)$ denotes the steering vector of the $i$-th microphone associated with the $l$-th plane wave. $ s_l(k) $ is the complex amplitude of the $l$-th plane wave, and $ n_i(k) $ is the noise received by the $i$-th microphone. The frequency domain received sound pressure model can be expressed in matrix form as:
\begin{equation}
p(k)=V(k, \Psi) s(k)+n(k),
\end{equation}
where $V(k, \Psi)$ is the $I \times L$ dimensional direction matrix. $s(k)=\left[s_1(k), s_2(k), \ldots, s_L(k)\right]^T$ is the $L$ dimensional source signal vector, $n(k)=\left[n_1(k), n_2(k), \ldots, n_I(k)\right]^T$ is the $I$ dimensional zero-mean Gaussian white noise vector, and $n(k)$ is assumed to be uncorrelated with $s(k)$. By N-point STFT, the $p(k)$ in the T-F domain can be recorded as $P_{i}(t, f)$:
\begin{equation}
P_{i}(t, f)=H_{i}(t, f) X_{i}(t, f)+V_{i}(t, f),
\end{equation}
where $t$ represents frame index and $f$ represents frequency bin obtained from STFT.
$X_{i}(t, f)$ and $V_{i}(t, f)$ represent the target and noise component, respectively. 
Considering the symmetry of $P_{i}(t, f)$ in frequency, $F = N/2 + 1$ is chosen throughout this paper.

\subsection{Spherical Harmonic Domain System Model}

In this section, we describe the proposed method for modeling acoustic signals in the spherical harmonic domain through the utilization of the SHT. 
By calculating the coefficients of spherical harmonics, the received speech signal at a specific point on the sphere surface can be estimated.
The spherical harmonics $Y_n^m(\theta, \phi)$ of order n $(n \in N)$ and degree m $(m \in Z$ and $-n \leq m \leq n) $ are defined as \cite{rafaely2015fundamentals}:
\begin{equation}
Y_n^m(\theta, \phi)=\sqrt{\frac{(2 n+1)}{4 \pi} \frac{(n-m) !}{(n+m) !}} P_n^m(\cos \theta) e^{i m \phi},
\end{equation}
where $(.)!$ is the factorial function, and $P_n^m$ is the normalized associated Legendre polynomial.
The spherical harmonic function $P_n^m(\cos \theta)$ captures the dependency on the elevation angle $\theta$, while the complex exponential term $e^{i m \phi}$ captures the dependency on the azimuth angle $\phi$.

According to the Fourier acoustic principle, the sampled sound pressure ${p}(k, r)$ and its spherical harmonic domain representation ${p}_{n m}(k, r)$ at frequency k and angle $(\theta, \phi)$ can be expressed as:
\begin{equation}
p(k, r)=\sum_{n=0}^{\infty} \sum_{m=-n}^n p_{n m}(k, r) Y_n^m(\theta, \phi).
\label{pkr}
\end{equation}

As explained in \cite{rafaely2015fundamentals}, the coefficients $p_{nm}$ diminish for kr in a range smaller than N and can therefore be neglected. Hence, Eq.\ref{pkr} can be approximated for an appropriate finite order N:
\begin{equation}
p(k, r) \cong \sum_{n=0}^N \sum_{m=-n}^n p_{n m}(k, r) Y_n^m(\theta, \phi),
\end{equation}
where N is the truncation order, $p(k, r)$ denotes the time-dependent amplitude of the sound pressure in free three-dimensional space, $p_{nm}(k,r)$ are the weights known as coefficients of the SHT, $k = 2\pi f/c$ is the wave number, $f$ is the frequency, and $c$ is the speed of sound in air. The coefficients $p_{nm}(k,r)$ are defined as \cite{rafaely2015fundamentals}:
\begin{equation}
p_{n m}(k, r)=\int_0^{2 \pi} \int_0^\pi p(k, \mathbf{r})\left[Y_n^m(\theta, \phi)\right]^* \sin (\theta) \mathrm{d} \theta \mathrm{d} \phi,
\end{equation}
where $(.)^*$ denotes complex conjugation. To satisfy the far-field condition, the distance $d$ between the sound source and the center of the microphone array must exceed $8r^{2}f/c$\cite{meyer2001beamforming}, where r is the array radius. This ensures negligible wavefront curvature effects. For $n \leq N$, $p_{nm}(k,r)$ can be obtained as:
\begin{equation}
p_{n m}(k, r) \cong \frac{4 \pi}{I} \sum_{i=1}^I p\left(k, \mathbf{r}_i\right)\left[Y_n^m\left(\theta_i, \phi_i\right)\right]^*,
\label{formula8}
\end{equation}
where $\mathbf{r}_i=\left(r, \theta_i, \phi_i\right)$ is the location of the $i$-th physical microphone and I is the number of physical microphones.

\section{Feature Extraction and Learning Framework}
\begin{figure}
    \centering
    \includegraphics[width=0.5\textwidth]{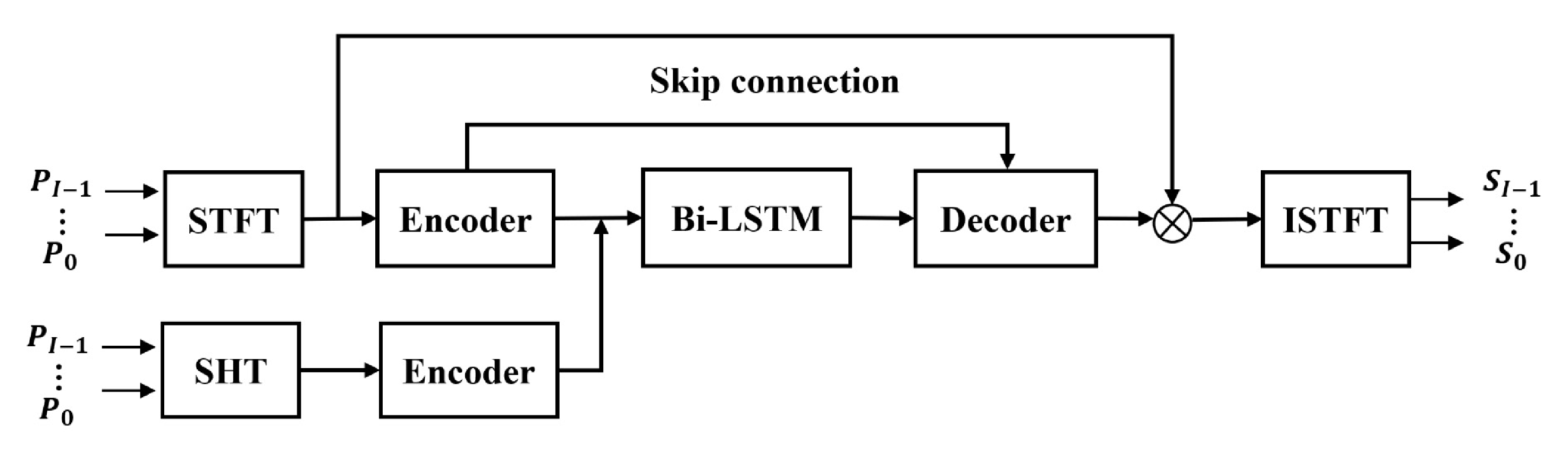}
    \caption{Defined spherical coordinate system.}
    \label{fig:fig2}
\end{figure}

We propose a novel architecture that fully leverages spatial information by using SHCs as auxiliary inputs to the model. This approach builds upon the baseline IGCRN model proposed in \cite{liu2021inplace}, which utilizes an encoder-decoder structure for multi-channel speech enhancement.
The key innovation is to introduce SHT for multi-channel speech enhancement to explore spatial clues, as illustrated in Fig. \ref{fig:fig2}.
In detail, the microphone array signals are transformed into the spherical harmonic domain to obtain SHCs, $p_{nm}(k,r)$, up to order $N$, which compactly encode the spatial distributions.
These coefficients form SHT are then provided as auxiliary inputs to a dedicated spatial encoder, along with the spectrograms from the STFT fed to the main spectro-temporal encoder (as in IGCRN).
Finally, the enhanced embeddings from both encoders are concatenated and passed to the decoder, which reconstructs the clean speech spectrogram.
Using SHT provides orientation-invariant coefficients that concisely capture useful spatial properties and inject global contextual information about the soundfield to guide the model.
The dual-encoder architecture enables joint modeling of spectral and spatial cues for improved speech enhancement.
The algorithm outlining the proposed system is presented in Algorithm \ref{algorithm}.

\begin{algorithm}[htb]
\caption{Algorithm for the proposed method. }
\label{alg:Framwork}
\begin{algorithmic}[1]
\REQUIRE ~~\\
A minibatch data $\{{\bf X}(t,f)_{mix},{\bf p}_{nm}(k,r)_{mix},{\bf s}(k)\}$.
Where ${\bf X}(t,f)_{mix}$ is the result of STFT of mixed speech. 
${\bf p}_{nm}(k,r)_{mix}$ denote the SHCs of order n and degree m, which are obtained by applying the SHT to the multi-channel mixed speech signal. 
$s(k)$ is the target speech.
learning rate is $\mu_d$.
\ENSURE ~~\\
The optimized proposed model
    \FOR{number of training iterations}
        \FOR{\textit{m}-th minibatch}

            \STATE $STFT_{out}=Encoder_{stft}(X(t,f)_{mix})$\
            \STATE $SHT_{out}=Encoder_{sht}(p_{nm}(k,r)_{mix})$\
            \STATE $LSTM_{out}=BiLSTM({STFT_{out},SHT_{out}})$\
            \STATE $est_{out}=Decoder(LSTM_{out})$
            \STATE $out=ISTFT(est_{out})$
            \STATE $Loss=MSE({out,s(k)})$
            \STATE $\theta_d\leftarrow \theta_d-\mu_d\frac{\partial Loss}{\partial\theta_d}$\
        \ENDFOR
    \ENDFOR
    \RETURN $\theta_d$
\end{algorithmic}
\label{algorithm}
\end{algorithm}

\section{EXPERIMENTS}
\subsection{Datasets and Evaluation Metrics}
The training data is generated by convolving multi-channel room impulse responses (RIRs) \cite{Alien1976Image} with speech signals from the TIMIT database \cite{zue1990speech}.
The TIMIT clips are divided into non-overlapping training, validation, and testing sets.
Noise clips from the DNS-Challenge corpus are used for training and validation, while NOISEX-92 \cite{varga1993assessment} and cafe noises from CHiME3 \cite{barker2015third} comprise the testing set.
RIRs are generated using the image method based on a 9-microphone uniform circular array with radius 0.035 m, randomly positioned inside a $6 \times 5 \times 4 \mathrm{~m}^{3}$ room. The source-array distance is 1 m.
RIRs are simulated with varying SNR (-6 to 6 dB) and RT60 (0.2 to 1 s) values.
For the purpose of evaluation, we define three distinct SNR levels: -5dB, 0dB and 5dB. In addition, we explore five distinct T60 values, ranging from 0.2s to 0.6s, with intervals of 0.1s. This comprehensive configuration results in the generation of 350 pairs for each specific case.

In this paper, perceptual evaluation of speech quality (PESQ)\cite{rix2001perceptual} and short-time objective intelligibility (STOI)\cite{taal2011algorithm} are chosen as the major objective metrics to evaluate the enhancement performance of different models.
PESQ rates speech quality on a scale from -0.5 to 4.5, while STOI gauges speech intelligibility on a scale of 0 to 100. Improved scores in both metrics reflect better performance.

\subsection{Experiment Setup}
\label{ssec:configurations}
\subsubsection{Network Detail}
The input features of the STFT and SHT are fed into two independent encoders. Each encoder consists of six cascaded $5 \times 1$ kernels inplace gated linear units (GLU), which are constructed using inplace convolutions as follows:
\begin{equation}
Y=E L U(B N(i \operatorname{Conv}(X) \otimes \operatorname{Sigmoid}(i \operatorname{Conv}(X)))),
\end{equation}
where $E L U(.)$ and $Sigmoid(.)$ are activation functions, $BN(.)$ is batch normalization, iConv is inplace convolution, and $\otimes$ denotes element-wise multiplication. To achieve a similar computational cost and number of parameters as IGCRN, which has 64 GLUs, we set the number of input channels for each GLU here to 32. The computational cost and number of parameters are shown in Table \ref{sec:Params}.
\begin{table}[t]
\centering
  \caption{Comparisoins of different approaches in Params and FLOPs.}
  \label{tab:hybrid}
    \begin{tabular}{@{}ccc@{}}
        \toprule
        Method & \#Params(M) & FLOPs(G) \\ \midrule
        IGCRN & 1.89 & 21.01 \\
        Proposed-parall & \textbf{1.82} & \textbf{19.52} \\ \bottomrule
    \end{tabular}
    \label{sec:Params}
\end{table}

After the encoders, we concatenate their outputs along the channel dimension and feed them into a channel-wise LSTM to refine the spatial information. The decoder consists of six cascaded inplace transpose GLUs with 128 input channels per transpose GLU.

\subsubsection{Training Detail}
For the SHT, N = 4, 25 spherical harmonics functions,
$Y_0^0(\theta, \phi)$,
$Y_1^{-1}(\theta, \phi)$,
$Y_1^0(\theta, \phi)$,
$Y_1^1(\theta, \phi)$,
$\cdots$,
$Y_4^4(\theta, \phi)$.
Then the complex value of each $Y_n^m\left(\theta_i, \phi_i\right)$ for the $i$-th microphone is specified. By employing (\ref{formula8}) a set of $p_{n m}(.)$ is calculated which is consist of 25 signals in the spherical harmonics domain.
All the utterances frame length is 32 ms and the frameshift 16 ms. The Square-root Hann window is used as the analysis window. The sampling rate is 16 kHz. A 512-point discrete Fourier transform is used to extract complex STFT spectrograms.
All models are trained using Adam optimizer with a fixed learning rate of 1e-3.
If validation loss does not decrease for consecutive two epochs, the learning rate will be halved. All models are trained for 60 epochs.

\subsection{Results and Discussions}
\label{ssec:Results}
\begin{table*}[t]
\centering
  \caption{PESQ results comparing proposed models with baselines.}
  \label{tab:table1}
  \resizebox{\linewidth}{!}{
        \begin{tabular}{ccccccccccccccccccc}
           \toprule
           \multirow{2}{*}[-2pt]{\textbf{Methods}} &
           \multicolumn{6}{c}{\textbf{-5dB}} & \multicolumn{6}{c}{\textbf{0dB}} & \multicolumn{6}{c}{\textbf{5dB}} \\
           \cmidrule(r){2-7} 
           \cmidrule(l){8-13}
           \cmidrule(l){14-19}
           & 0.2s & 0.3s & 0.4s & 0.5s & 0.6s & avg. & 0.2s & 0.3s & 0.4s & 0.5s & 0.6s & avg. & 0.2s & 0.3s & 0.4s & 0.5s & 0.6s & avg.\\
           \midrule
        Unprocessed
             & 1.36 & 1.33 & 1.31 & 1.29 & 1.29 & 1.32 & 1.52 & 1.48 & 1.44 & 1.39 & 1.40 & 1.45 & 1.83 & 1.76 & 1.70 & 1.62 & 1.61 & 1.70\\
        GCRN
             & 1.61 & 1.57 & 1.54 & 1.41 & 1.50 & 1.53 & 1.84 & 1.79 & 1.74 & 1.59 & 1.67 & 1.73 & 2.26 & 2.18 & 2.09 & 1.90 & 1.95 & 2.08\\
        IGCRN
             & 1.87 & 1.81 & 1.77 & 1.55 & 1.70 & 1.74 & 2.30 & 2.24 & 2.17 & 1.90 & 2.04 & 2.13 & 2.75 & 2.67 & 2.57 & 2.29 & 2.41 & 2.54\\
        \textbf{Proposed-serial}
             & \textbf{1.91} & \textbf{1.86} & \textbf{1.82} & \textbf{1.60} & \textbf{1.75} & \textbf{1.79} & \textbf{2.36} & \textbf{2.30} & \textbf{2.23} & \textbf{1.97} & \textbf{2.10} & \textbf{2.19} & \textbf{2.81} & \textbf{2.73} & \textbf{2.65} & \textbf{2.38} & \textbf{2.48} & \textbf{2.61} \\
        \midrule
        \textbf{Proposed-parallel}		
             & \textbf{2.19} & \textbf{2.03} & \textbf{1.85} & \textbf{1.65} & \textbf{1.75} & \textbf{1.89} & \textbf{2.66} & \textbf{2.51} & \textbf{2.30} & \textbf{2.06} & \textbf{2.13} & \textbf{2.32} & \textbf{3.10} & \textbf{2.94} & \textbf{2.72} & \textbf{2.48} & \textbf{2.51} & \textbf{2.75} \\
        \bottomrule
        \end{tabular}
    }
\end{table*}

\begin{table*}[t]
\large
\centering
  \caption{STOI results comparing proposed models with baselines.}
  \label{tab:table2}
  \resizebox{\linewidth}{!}{
        \begin{tabular}{ccccccccccccccccccc}
           \toprule
           \multirow{2}{*}[-2pt]{\textbf{Methods}} &
           \multicolumn{6}{c}{\textbf{-5dB}} & \multicolumn{6}{c}{\textbf{0dB}} & \multicolumn{6}{c}{\textbf{5dB}} \\
           \cmidrule(r){2-7} 
           \cmidrule(l){8-13}
           \cmidrule(l){14-19}
           & 0.2s & 0.3s & 0.4s & 0.5s & 0.6s & avg. & 0.2s & 0.3s & 0.4s & 0.5s & 0.6s & avg. & 0.2s & 0.3s & 0.4s & 0.5s & 0.6s & avg.\\
           \midrule
        Unprocessed
              & 54.95 & 52.00 & 50.09 & 43.59 & 46.93 & 49.51 & 64.57 & 62.24 & 59.83 & 53.24 & 55.85 & 59.15 & 76.08 & 73.90 & 71.14 & 65.19 & 66.29 & 70.52 \\
        GCRN
             & 62.29 & 59.66 & 57.85 & 50.29 & 54.82 & 56.98 & 71.71 & 69.25 & 67.36 & 61.28 & 64.44 & 66.81 & 81.18 & 78.92 & 77.03 & 72.76 & 74.03 & 76.78\\
        IGCRN
             & 71.09 & 68.68 & 66.91 & 59.51 & 63.66 & 65.97 & 80.79 & 78.84 & 76.99 & 71.68 & 74.03 & 76.47 & 88.08 & 86.40 & 84.66 & 81.15 & 82.22 & 84.50 \\
        \textbf{Proposed-serial}
             & \textbf{71.84} & \textbf{69.65} & \textbf{67.86} & \textbf{60.84} & \textbf{64.82} & \textbf{67.00} & \textbf{81.14} & \textbf{79.36} & \textbf{77.53} & \textbf{72.75} & \textbf{74.80} & \textbf{77.12} & \textbf{88.37} & \textbf{86.73} & \textbf{85.20} & \textbf{81.89} & \textbf{82.88} & \textbf{85.01} \\
        \midrule
        \textbf{Proposed-parallel}		
             & \textbf{77.73} & \textbf{75.06} & \textbf{71.24} & \textbf{65.63} & \textbf{67.39} & \textbf{71.41} & \textbf{84.66} & \textbf{82.69} & \textbf{79.60} & \textbf{75.74} & \textbf{76.34} & \textbf{79.81} & \textbf{90.27} & \textbf{88.47} & \textbf{86.28} & \textbf{83.51} & \textbf{83.59} & \textbf{86.42} \\
        \bottomrule
        \end{tabular}
    }
\end{table*}

\subsubsection{Performance for Proposed Structure}
We compare against two baseline models: GCRN, which uses convolutional recurrent networks for complex spectral mapping, and IGCRN, which extends GCRN with inplace convolutions.
We propose two variants of IGCRN:
\begin{itemize}
\item \textbf{Proposed-serial:} The SHT and STFT features are concatenated along the channel dimension and serially fed into a single IGCRN model.
\item \textbf{Proposed-parallel:} The SHT and STFT features are fed in parallel into two separate encoder branches of a dual-encoder IGCRN.
\end{itemize}
In Proposed-serial, the spatial and spectral features are combined into a single stream input to IGCRN. In Proposed-parallel, the SHT and STFT features are processed independently in dual encoder pathways before fusion. Both architectures augment the baseline with additional spherical harmonic spatial cues to enhance separation performance.

Experiments were conducted at -5dB, 0dB, and 5dB SNR levels, with 0.2s to 0.6s reverberation times. As shown in Tables \ref{tab:table1} and \ref{tab:table2}, the results demonstrate the superiority of the two proposed models utilizing SHT over the baseline model without SHT. Specifically, at -5dB SNR, Proposed-parallel achieved an average PESQ score of 1.89, while Proposed-serial scored 1.79, compared to 1.74 for the baseline IGCRN. The STOI results followed a similar trend, with both proposed models surpassing the baseline STOI score. Overall, the results indicate that the proposed models can effectively incorporate and leverage the spatial cues from the SHT to achieve marked improvements in speech quality and intelligibility over the baseline model. By capturing the spatial information in the multi-channel input via spherical harmonics, the proposed models significantly outperform the baseline lacking this capability.

\subsubsection{Ablation Study}
We further analyze the performance difference between our Proposed-serial and Proposed-parallel models. 
At an SNR of -5dB, the mean gains observed are 0.1 for PESQ and 4.41\% for STOI. As the SNR increases to 0dB, more substantial improvements are attained, with gains of 0.13 in PESQ and 2.69\% in STOI. Finally, at the highest tested SNR level of 5dB, the enhancements over the unprocessed signals are 0.14 for PESQ and 1.41\% for STOI, as shown in Tables \ref{tab:table1} and \ref{tab:table2}.
The gains from parallel encoders remain consistent as we vary the noise and reverberation levels. This indicates that modeling the SHT and STFT features separately enables better representations to be learned, compared to serially processing the concatenated features. The network can train more specialized encoders when the inputs are independent.
In contrast, the serial design forces the model to process the entire concatenated input in one encoder. This makes disentangling the spatial and spectrotemporal characteristics more challenging. The parallel approach does not have this constraint, allowing more robust joint representations to be formed after encoding.
In summary, our ablation study demonstrates the superior performance achieved by modeling the SHT and STFT features in parallel encoders rather than serially. The results clearly validate the benefits of the parallel architecture for multi-channel speech enhancement.

\section{CONCLUSION}
\label{sec:print}
In this work, we propose a method that fully leverages spatial information by using SHT as auxiliary model inputs. Spherical harmonics provide a compact representation that captures spatial cues across microphones. Experiments demonstrate that our model outperforms established benchmarks, remarkably with fewer computations and parameters. By leveraging spherical harmonics to incorporate directional cues, our model efficiently improves multi-channel speech enhancement performance.
An ablation study validates that superior performance is achieved by modeling the SHT and STFT features in parallel encoders rather than sequentially. This highlights the benefits of joint spatial-spectral modeling.
For future work, we intend to explore other applications of spherical harmonics for spatial audio and speech processing. The use of spherical harmonic coefficients as a form of spatial feature representation shows promising results for multi-channel speech enhancement.

\section{Acknowledgements}
This work was partly supported by the China National Nature Science Foundation (No. 61876214, No. 61866030).

\newpage

\bibliographystyle{IEEEbib}
\bibliography{strings,refs}

\end{document}